\documentclass[12pt]{article}
\usepackage{fancyhdr}

\usepackage{mathrsfs}
\usepackage[T1]{fontenc}
\usepackage{mathpazo}
\usepackage{setspace}
\usepackage{amsfonts}
\usepackage{amssymb}
\usepackage{amsmath}
\usepackage{epsfig}
\usepackage{latexsym}
\usepackage{color}
\usepackage{graphicx}
\usepackage{nicefrac}
\usepackage[latin1]{inputenc}
\usepackage{slashed}
\usepackage{multirow}
\usepackage{comment}
\usepackage{soul}
\usepackage{hyperref}
\usepackage[nosort]{cite}
\usepackage{datetime}

\usepackage[titletoc]{appendix}

\def\hybrid{\topmargin -20pt    \oddsidemargin 0pt
        \headheight 0pt \headsep 0pt
        \textwidth 6.25in       
        \textheight 9.25in       
        \marginparwidth .875in
        \parskip 5pt plus 1pt   \jot = 1.5ex}

\hybrid

\def\baselinestretch{1.2}

\catcode`\@=11

\def\marginnote#1{}
\newcount\hour
\newcount\minute
\newtoks\amorpm
\hour=\time\divide\hour by60
\minute=\time{\multiply\hour by60 \global\advance\minute by-\hour}
\edef\standardtime{{\ifnum\hour<12 \global\amorpm={am}%
        \else\global\amorpm={pm}\advance\hour by-12 \fi
        \ifnum\hour=0 \hour=12 \fi
        \number\hour:\ifnum\minute<10 0\fi\number\minute\the\amorpm}}
\edef\militarytime{\number\hour:\ifnum\minute<10 0\fi\number\minute}

\def\draftlabel#1{{\@bsphack\if@filesw {\let\thepage\relax
   \xdef\@gtempa{\write\@auxout{\string
      \newlabel{#1}{{\@currentlabel}{\thepage}}}}}\@gtempa
   \if@nobreak \ifvmode\nobreak\fi\fi\fi\@esphack}
        \gdef\@eqnlabel{#1}}
\def\@eqnlabel{}
\def\@vacuum{}
\def\draftmarginnote#1{\marginpar{\raggedright\scriptsize\tt#1}}

\def\draft{\oddsidemargin -.5truein
        \def\@oddfoot{\sl preliminary draft \hfil
        \rm\thepage\hfil\sl\today\quad\militarytime}
        \let\@evenfoot\@oddfoot \overfullrule 3pt
        \let\label=\draftlabel
        \let\marginnote=\draftmarginnote
   \def\@eqnnum{(\theequation)\rlap{\kern\marginparsep\tt\@eqnlabel}%
\global\let\@eqnlabel\@vacuum}  }

\def\preprint{\twocolumn\sloppy\flushbottom\parindent 2em
        \leftmargini 2em\leftmarginv .5em\leftmarginvi .5em
        \oddsidemargin -.5in    \evensidemargin -.5in
        \columnsep .4in \footheight 0pt
        \textwidth 10.in        \topmargin  -.4in
        \headheight 12pt \topskip .4in
        \textheight 6.9in \footskip 0pt
        \def\@oddhead{\thepage\hfil\addtocounter{page}{1}\thepage}
        \let\@evenhead\@oddhead \def\@oddfoot{} \def\@evenfoot{} }

\def\numberbysection{\@addtoreset{equation}{section}
        \def\theequation{\thesection.\arabic{equation}}}

\def\underline#1{\relax\ifmmode\@@underline#1\else
        $\@@underline{\hbox{#1}}$\relax\fi}

\def\titlepage{\@restonecolfalse\if@twocolumn\@restonecoltrue\onecolumn
     \else \newpage \fi \thispagestyle{empty}\c@page\z@
        \def\thefootnote{\fnsymbol{footnote}} }

\def\endtitlepage{\if@restonecol\twocolumn \else \newpage \fi
        \def\thefootnote{\arabic{footnote}}
        \setcounter{footnote}{0}}  

\catcode`@=12
\relax

\def\figcap{\section*{Figure Captions\markboth
        {FIGURECAPTIONS}{FIGURECAPTIONS}}\list
        {Figure \arabic{enumi}:\hfill}{\settowidth\labelwidth{Figure
999:}
        \leftmargin\labelwidth
        \advance\leftmargin\labelsep\usecounter{enumi}}}
 \relax
\def\tablecap{\section*{Table Captions\markboth
        {TABLECAPTIONS}{TABLECAPTIONS}}\list
        {Table \arabic{enumi}:\hfill}{\settowidth\labelwidth{Table
999:}
        \leftmargin\labelwidth
        \advance\leftmargin\labelsep\usecounter{enumi}}}
 \relax
\def\reflist{\section*{References\markboth
        {REFLIST}{REFLIST}}\list
        {[\arabic{enumi}]\hfill}{\settowidth\labelwidth{[999]}
        \leftmargin\labelwidth
        \advance\leftmargin\labelsep\usecounter{enumi}}}
 \relax
\makeatletter
\newcounter{pubctr}
\def\publist{\@ifnextchar[{\@publist}{\@@publist}}
\def\@publist[#1]{\list
        {[\arabic{pubctr}]\hfill}{\settowidth\labelwidth{[999]}
        \leftmargin\labelwidth
        \advance\leftmargin\labelsep
        \@nmbrlisttrue\def\@listctr{pubctr}
        \setcounter{pubctr}{#1}\addtocounter{pubctr}{-1}}}
\def\@@publist{\list
        {[\arabic{pubctr}]\hfill}{\settowidth\labelwidth{[999]}
        \leftmargin\labelwidth
        \advance\leftmargin\labelsep
        \@nmbrlisttrue\def\@listctr{pubctr}}}
 \relax
\makeatother
\newskip\humongous \humongous=0pt plus 1000pt minus 1000pt

\newif\ifdtup

\relax

\def\be{\begin{equation}}
\def\ee{\end{equation}}
\def\ba{\begin{eqnarray}}
\def\ea{\end{eqnarray}}

\def\del{\partial}

\def\A{\Alpha}
\def\b{\beta}

\def\g{\gamma}

\def\d{\delta}
\def\D{\Delta}

\def\th{\theta}

\def\m{\mu}

\def\om{\omega}

\def\l{\lambda}
\def\L{\Lambda}
\def\s{\sigma}

\def\cL{{\cal L}}

\def\no{\noindent}

\def\qq{\qquad}

\def\IR{\relax{\rm I\kern-.18em R}}

\def \A { {\bar A} }

\def \ha {{1\over 2}}

\def \ov {\over}

\def\IR{\relax{\rm I\kern-.18em R}}
\def\IL{\relax{\rm I\kern-.18em L}}

\def\inv{^{\raise.15ex\hbox{${\scriptscriptstyle -}$}\kern-.05em 1}}

\def\cL{{\cal L}}

\def\Tr{{\rm Tr}}

\begin{document}

\renewcommand{\theequation}{\thesection.\arabic{equation}}
\csname @addtoreset\endcsname{equation}{section}

\newcommand{\beq}{\begin{equation}}
\newcommand{\eeq}[1]{\label{#1}\end{equation}}
\newcommand{\ber}{\begin{equation}}
\newcommand{\eer}[1]{\label{#1}\end{equation}}
\newcommand{\eqn}[1]{(\ref{#1})}
\begin{titlepage}
\begin{center}


${}$
\vskip .2 in

{\large\bf Integrable Lorentz-breaking deformations and RG flows }

\vskip 0.4in

{\bf George Georgiou}\, and\, {\bf Konstantinos Sfetsos}
\vskip 0.15in

 {\em
Department of Nuclear and Particle Physics,\\
Faculty of Physics, National and Kapodistrian University of Athens,\\
Athens 15784, Greece\\
}

\vskip 0.12in

{\footnotesize \texttt {ggeo@phys.uoa.gr, ksfetsos@phys.uoa.gr}}


\vskip .5in
\end{center}

\centerline{\bf Abstract}

\no
We construct the all loop effective action for WZW models perturbed by current-bilinear terms of the type $J_+J_- $,
$J_+J_+ $ and $J_-J_- $, the last two of which explicitly break Lorentz invariance. For isotropic couplings
we prove integrability. For the case in which only the first two terms are present we identify a non-perturbative symmetry of the effective action and we compute the exact beta-functions.
These become identically zero outside a bounded region in the parametric space.

\vskip .4in
\noindent
\end{titlepage}
\vfill
\eject

\newpage

\tableofcontents

\noindent

\def\baselinestretch{1.2}
\baselineskip 20 pt
\noindent


\setcounter{equation}{0}
\section{Introduction }
A plethora of physical systems, especially in condensed matter physics, do not enjoy covariance under Lorentz transformations. This does not come as a surprise since the corresponding Lagrangians are effective ones aiming at a description of physics at low energies. In certain cases a group that contains the Lorentz group is replaced by some other group, e.g. instead of having the conformal group one has the Schrodinger group, which generically have less generators.  As a result symmetry constraints on the observables of the theory are less severe.

In all of these cases one faces the usual problem encountered in quantum field theory (QFT), namely that calculations can only be performed as a series expansion in some small parameter, thus rendering the description of strongly coupled systems
as intractable. It is the aim of this work to construct a class of two-dimensional field theories in which the Lorentz symmetry is broken in a controlled way so that it is possible to extract information about the exact in the couplings behavior of the theory.
The starting point will be a Lorentz invariant theory, namely the WZW model based on some semi-simple group $G$. We will perturb the theory with all possible terms bilinear in the currents, that is $J \bar J$, $JJ$ and  $\bar J \bar J$ (in the Euclidean
regime). It is apparent that the last two operators explicitly break Lorentz invariance. By using the gauging procedure used extensively in similar constructions \cite{Sfetsos:2013wia,Georgiou:2016urf,Georgiou:2017oly,Georgiou:2017jfi,Sfetsos:2017sep,Georgiou:2018hpd,Georgiou:2018gpe},
we will be able to find the all loop effective action of our models. From this, we will subsequently evaluate the exact in the deformation parameters $\beta$-functions of the theory by employing a variant of the background field method which in the context of $\l$-deformations was initially used in \cite{Appadu:2015nfa} and further exploited and extended in
\cite{Georgiou:2017aei,Georgiou:2017jfi,Georgiou:2018gpe,Sagkrioti:2018rwg}. The theory possesses a line of fixed points in the UV. We expect this situation to change in the case where two or more  WZW models at different levels are used, as it happened with the Lorentz preserving deformations of \cite{LeClair:2001yp,Georgiou:2016zyo,Georgiou:2017jfi,Georgiou:2018hpd,Georgiou:2018gpe}.

The second key feature of our models is that the deformations preserve integrability for all values of the deformations parameters. This is explicitly shown by establishing that the  equations of motion imply the existence of a Lax connection. Integrability of the theory is preserved in the entire flow and therefore these flows are called integrable.

Integrable Lorentz breaking deformations of two-dimensional conformal field theories (CFTs) have been recently studied in  \cite{Guica:2017lia}. In that case the CFT is deformed by an irrelevant operator which schematically takes the form
$J \bar T$, where $J$ is an Abelian  $U(1)$ current and $\bar T$ is the stress energy tensor. In a similar way to, the also irrelevant, $T \bar T$ deformation \cite{Smirnov:2016lqw,Cavaglia:2016oda}, the theory is solvable in the sense that its finite size spectrum can be obtained from that of the original CFT.
The holographic dictionary and a string theory realisation of $J \bar T$ deformations were studied in \cite{Bzowski:2018pcy} and \cite{Chakraborty:2018vja,Apolo:2018qpq} respectively, while the modular properties of the partition were analysed in \cite{Aharony:2018ics}.

Our construction differs from the aforementioned irrelevant deformations in several aspects. The main difference is that the deforming operators in our case have engineering dimension two and are, thus, classically marginal.
Furthermore, we can have complete control on the effective Lagrangian and we expect on the correlation functions of the theory as exact functions of the deformation parameters.  This can be achieved by using a method similar to
that used in similar Lorentz preserving constructions  \cite{Georgiou:2015nka,Georgiou:2016iom,Georgiou:2016zyo}.
The essence of the method relies on the construction of the corresponding all-loop effective actions
for the deformed theories which possess certain non-perturbative symmetries in the space of couplings
\cite{Itsios:2014lca,Sfetsos:2014jfa,Georgiou:2017jfi,Georgiou:2018hpd}.\footnote{ For isotropic $\l$-deformations the simplest such non-perturbative symmetry was found initially in \cite{Kutasov:1989aw} using path integral arguments.} One then can use low order perturbation theory in conjunction with the aforementioned non-perturbative symmetries to derive exact expressions for the observables of the theory.
In contrast, for the  $J \bar T$ deformed theories correlation functions were calculated to leading order in the deformation parameter in \cite{Guica:2019vnb} and no method to compute them exactly is apparently known.

\no
The plan of the present paper is as follows:
In section 2, we construct the all loop effective action and equations of motion of our models.
In section 3, we show that  the theory is classically integrable by finding the appropriate Lax connection.
In section 4, we consider a consistent truncation of our models that depends on two couplings and  identify the non-perturbative symmetries in the space of couplings.
In addition, we consider in this sector the non-Abelian T-duality and the pseudodual chiral limits.
In section 5, we calculate the exact in the deformation parameters $\b$-functions of our models which are particularly simple.
 In the last section we present our conclusions.

\section{The Lagrangian and the equations of motion}

In this section we construct the effective action for a certain class of a Lorentz non-invariant actions
and the corresponding equations of motion. Consider the WZW model action $S_{k}(g)$ at level $k$ for a
group element $g$ in a group $G$.
We add to this the action of a Lorentz-breaking principal chiral model (PCM)
constructed using the group element $\tilde g$ in the same group $G$. Namely,  we consider the action
\be
\label{oractt}
\begin{split}
& S_{k,E}(g,\tilde g)= S_{k}(g)
-{1\ov \pi}\int d^2\s\ \Big( \tilde g^{-1}\del_+\tilde g  \, E_{++} \, \tilde g^{-1} \del_+\tilde g + \tilde g^{-1}\del_-\tilde g \,  E_{--} \, \tilde g^{-1} \del_-\tilde g
\\
&\qq\qq\qq
+\tilde g^{-1}\del_+\tilde g \, E_{+-} \, \tilde g^{-1} \del_-\tilde g  \Big)\  ,
\end{split}
\ee
where the $E_{++}^{ab}, E_{--}^{ab}$ and $E_{+-}^{ab}$ are generic coupling matrices.
The first two can be taken to be symmetric with no loss of generality.
As in \cite{Sfetsos:2013wia,Georgiou:2016urf} we gauge the global
symmetry acting as
$g\to \L^{-1} g\L$ and $\tilde g\to \L^{-1} \tilde g $. Consider the
gauge invariant action
\be
\label{gauacc}
\begin{split}
& S_{k,E}(g,\tilde g, A_\pm)  = S_{k}(g,A_\pm)
-{1\ov \pi}\int d^2\s\ \Big( \tilde g^{-1}D_+\tilde g  \, E_{++} \, \tilde g^{-1} D_+\tilde g
\\
&\qq\qq + \tilde g^{-1}D_-\tilde g \,  E_{--} \, \tilde g^{-1} D_-\tilde g
+\tilde g^{-1}D_+\tilde g \, E_{+-} \, \tilde g^{-1} D_-\tilde g  \Big)\ ,
\end{split}
\ee
where the standard gauged WZW action is
\be
\begin{split}
&  S_{k}(g,A_\pm) = S_{k}(g)
+{k\ov \pi} \int d^2\s \ \Tr \big(A_- \del_+ g g^{-1}   - A_+ g^{-1} \del_- g
\\
& \qq\qq\qq + A_- g A_+ g^{-1}-A_-A_+\big)\ .
\end{split}
\ee
We have defined the covariant derivatives as
$D_\pm \tilde g= (\del_\pm -A_\pm) \tilde g$.
Next, we gauge fix in \eqn{gauacc} by setting $\tilde g=\mathbb{1}$. The one arrives at the
following action
\be
\begin{split}
& S_{k,\l_i}(g, A_\pm) = S_{k}(g)
+{k\ov \pi} \int d^2\s \ \Tr \big(A_- \del_+ g g^{-1}   - A_+ g^{-1} \del_- g+ A_- g A_+ g^{-1}\big) \\
&- {k \ov \pi} \int d^2\s\ \Big(\ha  A_+ \l_1 A_+  + \ha A_- \l_2 A_-
 +A_+ \l^{-1} A_-\Big)\ ,
\label{gaufix}
\end{split}
\ee
where we have redefined the PCM model coupling matrices as
\be
\begin{split}
 \l_1=2 {E_{++} \ov k} \ ,\qq \l_2= 2 {E_{--} \ov k} \ , \qq \l^{-1}= 1+{E_{+-} \ov k}\ .
\end{split}
\ee
The $\s$-model in obtained by integrating out the gauge fields since they appear in the action
only quadratically. To do that we use their equations of motion which are presented
later  in \eqn{dggd} in connection with integrability. In this way we find that
\be
\label{apm}
\begin{split}
 \left(\!\! \begin{array}{c}
    A_{+} \\ A_{-} \end{array}\!\!\! \right)= i \,  M^{-1}  \left(\!\! \begin{array}{c}
    -J_{-} \\ J_{+} \end{array}\!\!\! \right)\, ,
\end{split}
\ee
where the matrix $M$ is given
\be
\label{Mhat}
\begin{split}
M=
\left(  \begin{array}{cc}
      \l_1 &  \l^{-1}-D^T  \\
  \l^{-T}-D &  \l_2\\
  \end{array} \right)\ ,
\end{split}
\ee
with its inverse having entries
\be
\label{apmM}
\begin{split}
& M^{-1}_{11}=\Big( \l_1- (\l^{-1}-D^T)\l_2^{-1}(\l^{-T}-D)\Big )^{-1}\, ,
\\
&M^{-1}_{12}= \Big(1- (\l^{-T}-D)^{-1}\l_2(\l^{-1}-D^T)^{-1} \l_1\Big )^{-1}(\l^{-T}-D)^{-1}\, ,
\\
& M^{-1}_{21}= \Big(1- (\l^{-1}-D^T)^{-1}\l_1(\l^{-T}-D)^{-1} \l_2\Big )^{-1}(\l^{-1}-D^T)^{-1}\, ,
\\
& M^{-1}_{22}=\Big(\l_2- (\l^{-T}-D)\l_1^{-1}(\l^{-1}-D^T)\Big )^{-1}\, ,
\end{split}
\ee
which are themselves matrices.
The matrices $D_{ab}$ and the currents $J^a_{\pm}$ are defined as
\be
\label{hg3}
J^a_+ = - i\, \Tr(t^a \del_+ g g^{-1}) ,\qq J^a_- = - i\, \Tr(t^a g^{-1}\del_- g )\ ,
\qq D_{ab}= \Tr(t_a g t_b g^{-1})\  ,
\ee
where the $t^a$'s are Hermitian matrices. They obey $[t_a,t_b]=i f_{ab}{}^c t_c$, for some
real algebra structure constants and are normalized to one.

\no
Substituting the expressions for the gauge fields into the action \eqn{gaufix} gives the following
$\s$-model action
\be
\begin{split}
&  S_{k,\l}(g) = S_{k}(g)
 - {k\ov 2 \pi} \int  d^2\s  \left(\!\! \begin{array}{cc}
    -J_{-}\! &\! J_{+} \end{array}\!\!  \right)
M^{-1}  \left(\!\! \begin{array}{c}
    -J_{-} \\ J_{+} \end{array}\!\!\! \right) \, .
\end{split}
\label{defactigen}
\ee
This action is parity invariant under the interchange
\be
\label{pparri}
\s^+\leftrightarrow \s^- \ ,\qq   \l_1\leftrightarrow  \l_2\ ,\qq g\to g^{-1}\ .
\ee
We will see later that \eqn{defactigen}, at least for a truncated version, has a non-perturbative symmetry as well.

\section{Integrability}

In this section we  will demonstrate that the theory described by
\eqn{defactigen} is integrable for the case of isotropic couplings.

\no
Varying \eqn{gaufix} with respect
to $A_\pm$ and $B_\pm$ we find the constraints
\be
\label{dggd}
\begin{split}
&
D_+ g\, g^{-1} =  (\l^{-T}-\mathbb{1}) A_+ +   \l_2 A_- \ ,
\\
&
g^{-1} D_- g = -  \l_1 A_+ - (\l^{-1}-\mathbb{1})  A_-\ ,
\end{split}
\ee
where the covariant derivatives are defined as $D_\pm g= \del_\pm g -[A_\pm,g]$.
Solving for the gauge fields we obtain \eqn{apm} presented above.
Varying the action with respect to $g$ results into
\be
\label{eqg1g2}
D_ -(D_+ g g^{-1})= F_{+-} \quad \Longleftrightarrow\quad D_+(g^{-1}D_- g )
= F_{+-}\ ,
\ee
where the field strenghts are defined as usual
\be
F_{+-}=\del_+ A_- - \del_- A_+ - [A_+,A_-]\ .
\ee
Substituting the constraints \eqn{dggd} in \eqn{eqg1g2} one obtains
\be
\begin{split}
\label{eomAinitial1}
& \l_1\del_+ A_+ +\l^{-1} \del_+  A_- -\del_-A_+
= [ A_+,\l^{-1} A_-] + [A_+, \l_1 A_+] \ ,
\\
& \l_2\del_- A_- +\l^{-T} \del_-  A_+ -\del_+A_-
=-  [ \l^{-T} A_+,A_-] + [A_-, \l_2 A_-] \ .
\end{split}
\ee
These are written solely in terms of the gauge fields.  The group elements are implicitly present
via \eqn{apm}.

\no
We restrict to the case in which all coupling
matrices are proportional to the identity, using however the same symbol for the
proportionality constant, i.e.
\be
\label{ddiaa}
( \l_i)_{ab} =  \l_i \d_{ab}\ ,\quad  i=1,2\ , \quad \l_{ab}=\l\d_{ab}\ .
 \ee
 Note that, in this case the last brackets in \eqn{eomAinitial1} vanish.
One can now make the following ansatz for the Lax connection, namely
\be\label{Lax-ansatz}
\cL_+=c_1 A_++c_2 A_- \ ,\qq \cL_-=c_3 A_+ +c_4 A_-\ .
\ee
The constants $c_i, \, i=1,2,3,4$ may depend on the coupling constants of the theory, as well as on the spectral parameter $z$.
Note that, the form of the ansatz does not respect Lorentz invariance in accordance with that of the action.
Substituting \eqn{Lax-ansatz} in the Lax equation
\be
\label{Lax-gen}
\partial _+\cL_--\partial _-\cL_+ -[\cL_+,\cL_-]=0
\ee
 and plugging in the resulting equation the expressions for $\del_+ A_+$ and $\del_- A_-$ that can be obtained from \eqn{eomAinitial1}
we get the following set of equations by demanding that the coefficients of $\del_+  A_-$, $\del_-  A_+$ and $[ A_+,A_-]$ vanish
\be\label{set-Lax}
\begin{split}
c_3 a_1+c_4-c_2 a_2=0\ ,\quad c_3 b_1-c_1-c_2 b_2=0\ ,\quad -c_3 a_1-c_2 b_2+c_2 c_3-c_1 c_4=0\ ,
\end{split}
\ee
where
\be\label{set-Lax}
\begin{split}
& a_1=- \l^{-1} \l_1^{-1},\qq a_2= \l_2^{-1},\qq b_1=\l_1^{-1},\qq b_2=- \l^{-1} \l_2^{-1}\ .
\end{split}
\ee
Hence, we have three equations for four unknowns. In particular,
one may identify $c_2\equiv z$ with the spectral parameter $z$ and solve the set of three equations in \eqn{set-Lax} for $c_1$, $c_3$ and $c_4$ to get
\be\label{Lax-sol}
\begin{split}
&c_1= \frac{\sqrt{C}+ (a_2 b_1 -a_1 b_2-1)z+a_1}{2 a_1}, \qq
c_3= \frac{\sqrt{C}+ (a_1 b_2+a_2 b_1-1)z+a_1}{2 a_1 b_1} \\
& c_4= -\frac{\sqrt{C} +(a_1 b_2 -a_2 b_1 -1)z+a_1}{2 b_1}\ , \\
& C=(a_1+z (a_1b_2+a_2 b_1-1))^2+4 a_1 b_1 b_2 z (1-a_2  z)\ .
\end{split}
\ee
These values for the coefficients of the Lax connection quarrantee that the  flatness condition for the Lax pair is satisfied for all values of the spectral parameter $z$.

\section{ Truncation to a two-parameter integrable model}

In  this section we discuss in detail a two-parameter model which arises by taking the limit $ \l_2 \to 0$.
This is a consistent truncation of the full theory even at the quantum level as we will see below in the calculations of the $\b$-functions of the model.

\subsection{Truncation of the action and the equations of motion}

In the limit $\l_2\to 0 $ and after redefining the coupling matrix $\l_1$ to a new one as $\tilde\l$ as
\be
\label{redff}
 \l_1 = \l^{-1}\tilde\l \l^{-T}\ ,
\ee
the action \eqn{defactigen} becomes
\be
\begin{split}
&  S_{k,\l}(g) = S_{k}(g) +{k\ov \pi}  \int  d^2\s\ \Big[ J_{+} (\l^{-1} - D^T)^{-1} J_{-}
\\
&\phantom{xxxxxxxxxx}  +\ha J_+  (\mathbb{1} -\l D^T)^{-1}\tilde \l (\mathbb{1} - D\l^T)^{-1} J_+\Big] \ .
\end{split}
\label{dlimit}
\ee
The first line is the original $\l$-deformed model action \cite{Sfetsos:2013wia},
whereas the second one represents the Lorentz-breaking term. Note that, the parameter $\tilde \l$, unlike $\l$, appears linearly.
The equations of motion for this two-parameter $\s$-model case can be obtained by
taking the limit $\l_2 \to 0$ and using the redefinition \eqn{redff} in \eqn{eomAinitial1}.

\no
It can be readily checked that \eqn{dlimit} is invariant under the non-perturbative $Z_2$-symmetry
\be
\label{symmm}
\l\to \l^{-1}\ ,\qq \tilde \l \to - \lambda^{-1} \tilde \lambda \l^{-T}\ , \qq k\to -k\ ,\qq g\to g^{-1}\ .
\ee
This is a direct analog of the corresponding symmetries found in \cite{Itsios:2014lca} for the original single $\l$-deformed actions and their generalizations.
Due to  the Lorentz-breaking term, a boost transformation in the space of the two-dimensional coordinates
doesn't leave invariant the action, unless it is accompanied by a transformation of the parameter $\tilde \l$.
It is given by
\be
\label{boostsym}
\s^\pm \to e^{\pm v} \s^\pm\ ,\qq \tilde\l\to e^{2 v}\tilde\l \ ,
\ee
where $v$ is a constant.
Both of the above symmetries will be reflected in the expressions of the $\b$-functions as we will see in the next section.

\no
For small values of $\l$ and $\tilde \l$ the action \eqn{dlimit} expands as
\be
\begin{split}
& S_{k,\l}(g) = S_{k}(g) + {k\ov \pi}\int  d^2\s\
\Big( J_{+}\l J_{-} + \ha  J_{+} \tilde \l J_{+} \Big)+ {\cal O}(\l^2,\l\tilde \l) \ .
\end{split}
\label{dlimit3}
\ee
Hence the deformation from the original WZW model is driven by current bilinears.
The Lorentz-breaking term is, for isotropic $\tilde\l$, proportional to the energy momentum
tensor of the WZW CFT.
Thus, all loop effects in $\l$ and $\tilde \l$ are taken into account in \eqn{dlimit} which is the corresponding effective action. Small curvatures are guaranteed as long as $k\gg 1$.

We discuss next the conditions on the parametric space that are necessary for absence of singularities and having a metric of
 Euclidean signature. We do so by restricting to the case of diagonal couplings
 \be
 \l_{ab}=\l\d_{ab}\ ,\qq \tilde \l= \tilde \l \d_{ab}\ .
\ee
Since the matrix $D$ is orthogonal it has eigenvalues lying on the unit circle. Therefore,
to avoid singularities we should restrict to $|\l|<1$.
In addition, the Euclidean signature is guaranteed provided that the parameters $\l$ and $\tilde \l$ are such that they lie
outside the parabola, i.e.
\be
\label{bounds}
\tilde \l +1 > \l^2 \ .
\ee
Note that the symmetry \eqn{symmm} maps the inside to the outside of this parabola.

\subsection{The non-Abelian and pseudodual chiral limits}

There are two other interesting limits of the action \eqn{dlimit}.
In the first limit and similar to that in \cite{Sfetsos:2013wia} one expands the matrix and group elements near the identity as
\begin{equation}
\lambda_{ab} =\delta_{ab} - {E_{ab} \ov k}  +  \cdots\ , \qquad g = \mathbb{I} + i { v_a t^a \ov k} + \cdots  \ ,
\qq k\to \infty\ ,
\label{laborio}
\end{equation}
where $E$ is a general $\text{dim}G$ square matrix. This leads to
\begin{equation}
J_\pm^a = {\del_\pm v^{a}\ov k} + \cdots  \ ,\qquad
D_{ab} = \delta_{ab}+\frac{f_{ab}}{k} +  \cdots \ ,\qquad {  f_{ab} =  f_{abc} v^c}\ .
\label{orrio}
\end{equation}
In this limit and after the rescaling $\tilde \l\to \tilde \l/k$, the action \eqn{dlimit} becomes
\begin{equation}
\begin{split}
& S_{\rm non\!-\!Abel}(v) = {1\ov 2 \pi} \int \text{d}^2\s\, \Big[ \del_+ v (E +f)^{-1}\del_- v
\\
&\qq\phantom{xxxxxxxxxxxx} + \del_+ v (E+f)^{-1}\tilde \l (E^T-f)^{-1}\del_+ v \Big] \ .
\label{nobag}
\end{split}
\end{equation}
We note that in this limit the WZW part of the action vanishes and the Lorentz-breaking term has a limit on its own.
The above action
 is nothing but the non-Abelian T-dual with respect to the $G_{L}$ action of the $\sigma$-model given by  the PCM action
with general coupling matrix $E_{ab}$ plus a Lorentz-breaking term.

\no
In the second limit we take, as in \cite{Georgiou:2016iom}, that
\begin{equation}
\lambda_{ab}= -\delta_{ab} +{E_{ab} \ov k^{1/3}}+ \cdots \ ,\qq g= \mathbb{I} + i {v^a t^a\ov k^{1/3}}  + \cdots\ ,\qq k\to \infty\ ,
\label{limps}
\end{equation}
where again $E$ is a general $\text{dim}G$ square matrix. The various quantities expand as in \eqn{orrio} with $k$ replaced by $k^{1/3}$.
Then the action, after the rescaling $\tilde \l\to \tilde \l/k^{1/3}$,  becomes
\begin{equation}
S_{\rm pseudodual}={1\ov 8\pi} \int\text{d}^2\sigma\, \Big[\del_+v \Big(E +\frac{1}{3} f\Big)\del_-v
+ \del_+v \tilde \l \del_+v  \Big] \, .
\label{psac}
\end{equation}
The matrix $E$ can be taken to be symmetric since any antisymmetric piece leads to a total derivative.
The quadratic part of the WZW action and the Lorentz preserving deformation term  are combined for
the limit \eqn{limps} to exist, whereas as before the Lorentz violating term has a limit on its own.
This action \eqn{psac} is nothing but the pseudodual model action for PCM found in \cite{Georgiou:2016iom},
which generalized  the pseudodual model action of \cite{Nappi:1979ig} obtained for $E_{ab}\sim \delta_{ab}$,
plus a Lorentz violating term.

\section{Renormalization group flows}

In this section we compute the $\b$-function equations for the couplings.
For Lorentz-invariant $\s$-models one should in principle make use to the general equations involving the RG for two-dimensional
$\s$-models \cite{honer}. However, this is not possible for the models and hand,
so that we will use an alternative method initiated in \cite{Appadu:2015nfa}
for the isotropic case for $\l$-deformations and since it has been extended and applied in full
generality in \cite{Sagkrioti:2018rwg}. In all cases so far the $\s$-models were of course Lorentz-invariant, so that we cannot use existing results.

\subsection{The $\beta$-functions}

We choose a particular configuration of the group elements in order to compute the running of couplings. In particular,
$g = e^{\s^+ \th_+ + \s^- \th_-}$, where the  matrices  $\th_\pm$
are constant and commuting. Then we have that $J_{\pm}=-i \th_\pm$ and that the matrix $D=\mathbb{1}$.
Additionally, the expressions
\eqn{apm}  for the gauge fields become\footnote{
One might wonder to what extend the result obtained in this way is background independent.
The use of specific group elements is justified by the consistency of the end result. In addition, even when
using arbitrary  group elements as backgrounds, the end result is indeed background independent \cite{Sagkrioti:2018rwg}.}
\be
\label{gg23}
\left(  \begin{array}{c}
A^{(0)}_+\\
A^{(0)}_-\end{array} \right)={1\ov \D}
\left(  \begin{array}{cc}
     -\l^2  \l_2 &  \l(1-\l)  \\
   \l(1-\l)  &   - \l^2  \l_1 \\
  \end{array} \right)
  \left(  \begin{array}{c}
  -\th_-\\
\th_+\end{array} \right)\ ,
\ee
where superscript denotes the fact that these are classical values for the gauge fields and we have defined that
\be
\D = (1-\l)^2 -\l^2  \l_1 \l_2\ .
\ee
Then the Lagrangian density for the action \eqn{defactigen} reads
\be
\cL^{(0)}=-{k\ov 2 \pi} {1\ov \D} \Big( \l^2 \l_1 \th_+\th_+  +  \l^2 \l_2 \th_-\th_-
+ (1-\l^2\big(1+ \l_1\l_2)\big)  \th_+\th_-  \Big)\ .
\label{gg233}
\ee
The next step is to consider the fluctuations of the gauge fields around \eqn{gg23} and let
\be
A_\pm = A^{(0)}_\pm + \d A_\pm \ ,\qq
(\tilde A_\pm^{(0)})_{ab} =i f_{abc} (A_\pm^{(0)})_c\ .
\ee

\no
The classical equations of motion in the case of diagonal couplings become
\be
\begin{split}
&
\l \l_2 \del_- A_- + \del_-A_+ - \l \del_+ A_- + [A_+,A_-] = 0 \ ,
\\
&
\l \l_1 \del_+ A_+ + \del_+A_- - \l \del_- A_+ -  [A_+,A_-] = 0\ .
\end{split}
\ee
Note that the classical background solution given by \eqn{gg23} indeed solves these system.
Then, the linearized fluctuations for the classical equations of motion are
\be
\begin{split}
&
(\l \l_2 \del_- - \l \del_+ - \tilde \A_+^{(0)})\d A_- + (\del_-+  \tilde \A_-^{(0)}) \d A_+ = 0 \ ,
\\
&
(\l \l_1 \del_+- \l \del_- - \tilde \A_-^{(0)})\d A_+ + (\del_+ +  \tilde \A_+^{(0)}) \d A_- = 0\ .
\end{split}
\ee
These can be cast in the form
\be
\hat D \left(
         \begin{array}{c}
           \d A_-  \\
           \d A_+ \\
         \end{array}
       \right) = 0\ ,
\ee
where the operator $\hat D$ is first order in the worldsheet derivatives.
Its form in the Euclidean
regime and in momentum space will be presented following the conventions of \cite{Georgiou:2018hpd}.
This amounts to replacing $(\del_+,\del_-) \to \ha (\bar p,p)\equiv (p_+,p_-)$.
Then we have that $ \hat D = \hat C + \hat F $, where
\be
\label{chatt}
\hat C = \left(
           \begin{array}{cc}
             \l \l_2  p_- -\l p_+ &  p_-  \\
              p_+ &  \l \l_1  p_+ -\l p_-  \\
           \end{array}
         \right)
\ee
and
\be
\label{Fhatt}
\hat F =\left(
           \begin{array}{cc}
          - \tilde A_+^{(0)} &  \tilde A_-^{(0)} \\
        \tilde A_+^{(0)} & -\tilde A_-^{(0)} \\
           \end{array}
         \right)\  .
\ee
Integrating out the fluctuations, gives the effective Lagrangian of our models
\be
-\cL_{\rm eff} = \cL^{(0)} + \int^\m {d^2 p\ov (2\pi)^2} \ln (\det \hat D)^{-1/2}\ .
\ee
This integral is logarithmically divergent with respect to the UV mass scale
$\m$.  In order to isolate this we perform a large momentum expansion of the integrand.
We will keep terms proportional to $\displaystyle {1\ov |p|^2}$, where $|p|^2=p\bar p$.
Since $\hat C$ grows with $|p|$ we use the fact that
\be
\ln (\det \hat D)  = \ln \det \hat C + \Tr (C^{-1} \hat F) -\ha \Tr(C^{-1} \hat F)^2 + \cdots  \ .
\ee
The only term in the above expansion that contributes is the last one, obtaining
\be
\label{fhhreff}
-\cL_{\rm eff} = \cL^{(0)} + {1\ov 16 \pi^2} \int^\m d^2 p\, \Tr(C^{-1} \hat F)^2 + \cdots \ .
\ee
Next we use the fact that the integration measure is  $d^2 p = r dr d\phi$, that $p=re^{i\phi}$, $\bar p=re^{-i\phi}$
and evaluate $\Tr(C^{-1} \hat F)^2$. The dependence on $r$ is of the form $1/r^2$ which upon integration
gives the necessary factor of $\ln \m$.
Then
\be
\label{hgghr}
-\cL_{\rm eff} = \cL^{(0)} +{c_G  \ov 32 \pi^2} \ln \m^2
\Big(C^{+-} A^{(0)}_+ A^{(0)}_-  + C^{++} A^{(0)}_+ A^{(0)}_+  + C^{--} A^{(0)}_- A^{(0)}_- \Big) \ ,
\ee
were we used that $\Tr(\tilde A_+^{(0)} \tilde A_-^{(0)})= c_G (A_+^{(0)})^a
(A_-^{(0)})^a$, etc., where $c_G$ is the eigenvalue of the quadratic Casimir in the adjoint
representation defined as $f_{acd}f_{bcd}=c_G \d_{ab}$. The coefficients $C_{\pm\pm}$ and $C_{+-}$ are computed to be
\be
\label{cppm}
\begin{split}
& C^{+-}(\l, \l_1, \l_2)= 8\int_0^{2\pi}d\phi\, {1\ov Z^2}
\big(1-\l+\l \l_1 e^{ i \phi}\big) \big(1-\l+\l \l_2 e^{-i \phi}\big)\ ,
\\
& C^{++}(\l, \l_1, \l_2)=
4\int_0^{2\pi}d\phi\, {e^{-i\phi}\ov Z^2} \big(1-\l+\l \l_1 e^{ i \phi}\big)^2\ ,
\\
& C^{--}(\l, \l_1,\l_2) = 4\int_0^{2\pi}d\phi\, {e^{i\phi}\ov Z^2}
\big(1-\l+\l \l_2 e^{-i \phi}\big)^2\ ,
\end{split}
\ee
where
\be
Z =1-\l^2 (1+ \l_1 \l_2) + \l^2  \big(\l_1  e^{i\phi} + \l_2  e^{-i\phi}\big) \ .
\ee
The above integrals are non-trivial due to the non-vanishing coefficients $\l_1$ and $\l_2$ which are
responsible for the breaking of Lorentz invariance.  Due to \eqn{pparri} we have the identities
$C^{++}(\l,\l_1,\l_2)\!=\! C^{--}(\l,\l_2,\l_1)$ and $C^{+-}(\l, \l_1,\l_2)\!=\!
C^{+-}(\l,\l_2, \l_1)$.

\subsubsection{Restriction to the two-parameter model}

For simplicity we will restrict to the case of the  two-parameter model  which taken into account  the redefinition \eqn{redff}
means that
\be
 \l_2=0\ ,\qq  \l_1= \tilde \l/\l^2 \ .
\ee
Then the background gauge fields \eqn{gg23} become
\be
\label{gg23r}
\left(  \begin{array}{c}
A^{(0)}_+\\
A^{(0)}_-\end{array} \right)=
\left(  \begin{array}{cc}
    0 &  {\l\ov 1-\l}  \\
   {\l\ov 1-\l}  &   - {\tilde \l\ov (1-\l)^2} \\
  \end{array} \right)
  \left(  \begin{array}{c}
  -\th_-\\
\th_+\end{array} \right)\ .
\ee
Also the classical Lagrangian density corresponding to the action \eqn{gg233} simplifies to
\be
\cL^{(0)}=-{k\ov 2 \pi}  \Big( {1+\l\ov1-\l}  \th_+\th_- +{ \tilde \l \ov (1-\l)^2} \th_+\th_+  \Big)\ .
\label{gg233r}
\ee
The integrals \eqn{cppm} simplify as (redefinition of the arguments is implied)
\be
\label{cppm}
\begin{split}
& C^{+-}(\l,\tilde \l)= 8 {1-\l\ov \l} \int_0^{2\pi}d\phi\,
{\l(1-\l)+\tilde \l e^{i\phi}\ov (1-\l^2 +\tilde \l e^{i \phi})^2}\ ,
\\
& C^{++}(\l,\tilde \l)=
 {4\ov \l^2} \int_0^{2\pi}d\phi\, e^{-i\phi}
{(\l(1-\l)+\tilde \l e^{i\phi})^2\ov (1-\l^2 +\tilde \l e^{i \phi})^2}\ ,
\\
& C^{--}(\l,\tilde \l) = 4(1-\l)^2 \int_0^{2\pi}d\phi\,
{e^{i\phi}\ov (1-\l^2 +\tilde \l e^{i \phi})^2}\ ,
\end{split}
\ee
Moreover, the integrals can be transformed into a contour integral in the complex plane by letting as usual $z=e^{i\phi}$. They
read
\be
\label{cppmm}
\begin{split}
& C^{+-}(\l,\tilde \l)= - 8 i  {1-\l\ov \l} \oint_C dz \,
{\l(1-\l)+\tilde \l z\ov z(1-\l^2 +\tilde \l z)^2}\ ,
\\
& C^{++}(\l,\tilde \l)= -i
 {4\ov \l^2} \oint_C dz\,
{(\l(1-\l)+\tilde \l z)^2\ov z^2(1-\l^2 +\tilde \l z)^2}\ ,
\\
& C^{--}(\l,\tilde \l) =- 4i  (1-\l)^2 \oint_C dz\,
{1\ov (1-\l^2 +\tilde \l z)^2}\ ,
\end{split}
\ee
where $C$ is the unit circle centered at $z=0$ and transversed counter clockwise.
Consider the first two integrals. There are poles at $z=0$ and at $\displaystyle z=\tilde z$, where
\be
\tilde z \equiv {\l^2-1\ov \tilde \l} \ .
\ee
We note that this is invariant under \eqn{symmm} and can be either outside or inside $C$.
Considering the first case and using the residue corresponding to the pole
at $z=0$ we obtain that
\be
|\tilde z|>1: \qquad  C^{+-}(\l,\tilde \l)= {16\pi \ov (1+\l)^2}\ ,\qq C^{++}(\l,\tilde \l)
= {16\pi \tilde \l\ov \l(1-\l)(1+\l)^3}\ .
\ee
In the second case with $\tilde z$ in the interior of $C$, we obtain $C^{+-}\!\!=C^{++}\!\!=0$.
Finally,
\be
C^{--}(\l,\tilde \l)\!\!=0\ ,
\ee
identically, for either case. Using the above, and assuming that $|\tilde z|>1$,
the second term in \eqn{hgghr} becomes
\be
\label{hghr448}
\begin{split}
&
{c_G  \ov 32 \pi^2} \ln \m^2
\Big(C^{+-} A^{(0)}_+ A^{(0)}_-  + C^{++} A^{(0)}_+ A^{(0)}_+  + C^{--} A^{(0)}_- A^{(0)}_- \Big)
\\
& \qq= {c_G\ov 2\pi} \ln\m^2 \left({1\ov (1+\l)^2} A^{(0)}_+ A^{(0)}_- + {\tilde \l\ov \l(1-\l)(1+\l)^3} A^{(0)}_+ A^{(0)}_+\right)
\\
&\qq =- {c_G\ov 2\pi} \ln\m^2  \left({\l^2\ov (1-\l^2)^2} \th_+ \th_- + {\l^2 \tilde \l\ov (1-\l^2)^3} \th_+ \th_+\right)\ ,
\end{split}
\ee
where we have used \eqn{gg23r}.

\no
Next we demand that this action is $\m$-independent, i.e. $\del_{\ln \m^2} \cL_{\rm eff}=0$.
To leading order in $k$ this derivative acts only on the coupling constants in $\cL^{(0)}$ giving
\be
{d\ov d\ln \m^2} \cL^{(0)} = -{k\ov \pi} {\b_\l\ov (1-\l)^2 }\th_+\th_-
-{k\ov2 \pi} \left({ \b_{\tilde \l}\ov (1-\l)^2} + {2\tilde \l \b_{\l}\ov (1-\l)^3}\right) \th_+\th_+\ ,
\ee
which has precisely the same structure as \eqn{hghr448}. This observation is closely related to the fact that
truncating  the full theory to the one with two couplings is consistent with the RG equations.
Finally, imposing the condition $\del_{\ln \m^2} \cL_{\rm eff}=0$, we
get that the $\b$-functions are given by
\be
\label{systrg}
|\tilde z|>1: \qquad
\b_\l  = -{c_G\ov 2k} {\l^2\ov (1+\l)^2} \ ,
\qq
\b_{\tilde \l} ={c_G\ov k} {\l^3 \tilde \l\ov (1-\l) (1+\l)^3}\  .
\ee
We note that this system is indeed invariant under the symmetries \eqn{symmm} and \eqn{boostsym} of the effective action \eqn{dlimit}. In particular, the latter symmetry dictates the linear in $\tilde\l$ form of $\b_{\tilde \l}$.

\no
It can be checked that $\tilde z$ is a RG flow constant. Hence, $\b_\l$ suffices and
\be
\label{parabb}
\tilde\l = {\l^2-1\ov \tilde z}\ .
\ee
The RG flows are depicted in the Figure 1. The $(\l,\tilde \l)$-plane can be divided into regions I,II,II and IV defined as
\be
\begin{split}
 {\rm region\ I}:& \phantom{xxxxxx} 
 {\rm corresponding\ to}\quad \ -1<\tilde z <0\ ,
\\
 {\rm region\ II}:& \phantom{xxxxxx}  
  {\rm corresponding\ to}\quad \ \tilde z <-1\ ,
\\
 {\rm region\ III}:& \phantom{xxxxxx} 
  {\rm corresponding\ to}\quad \ \tilde z >1\ ,
 \\
 {\rm region\ IV}:& \phantom{xxxxxx} 
 {\rm corresponding\ to}\quad\  0<\tilde z <1\ .
 \end{split}
\ee
In regions I and IV the $\b$-functions vanish and there is no flow.
Since $k$ is a positive integer, the physical region is that with $-1<\l<1$. For this range of $\l$, the
region IV is not physical since in there the signature is negative, whereas in the regions I, II and III the signature is positive. In regions II and III the $\b$-functions are non-vanishing and are given by \eqn{systrg}.
Note that the transition from the conformal region I into II occurs precisely when the Lorentz violating second term in \eqn{gg233r} becomes smaller than the first term that
preserves Lorentz invariance. This has consequences for the stability of the theory as we will
readily see in the next subsection.
\begin{figure}[h] 
\label{betaplot4}
\begin{center}
\vskip -14 cm
\includegraphics[height= 23 cm,angle=0]{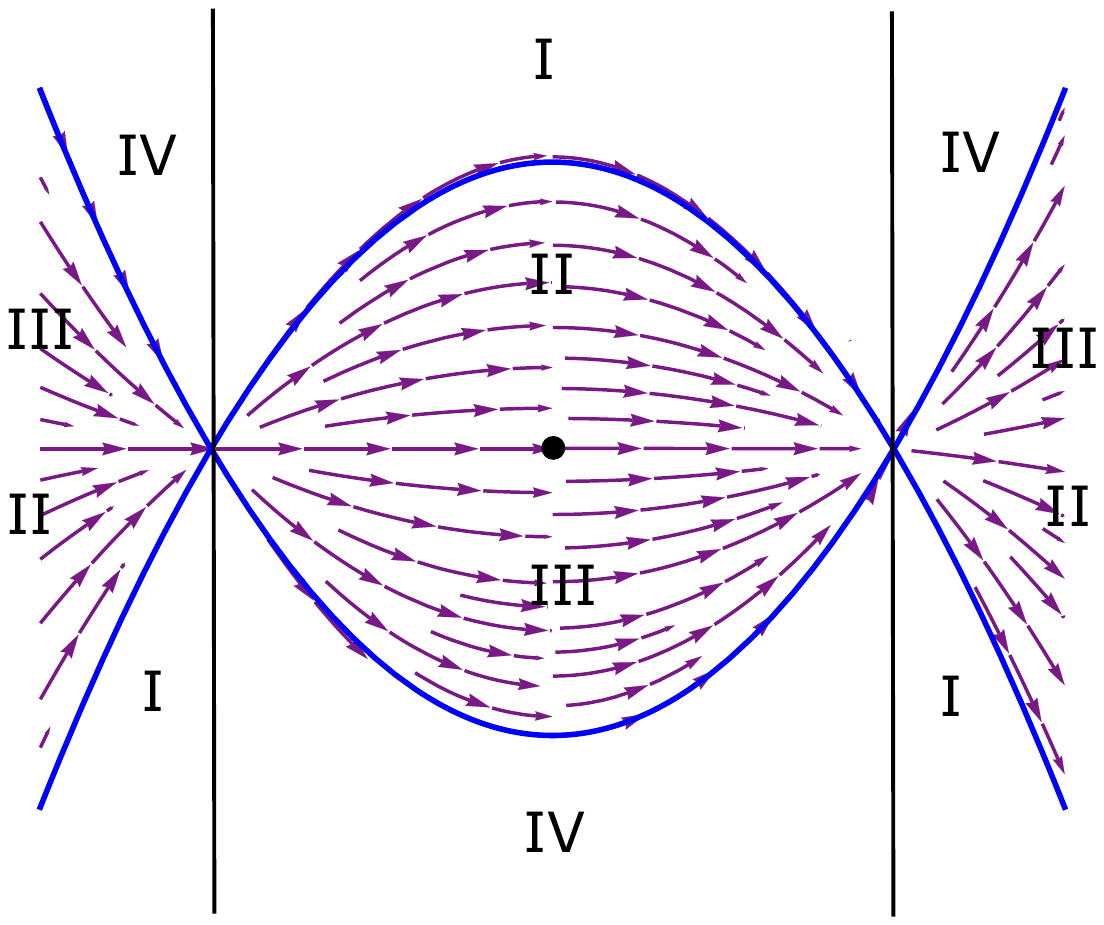}
\end{center}
\vskip -2 cm
\caption{RG flows in the $(\l,\tilde \l)$-plane ($\l$-horizontal) with $k=c_G=1$.
The flows are parabolas and the
blue curves, from top down, are those with $\tilde \l=1-\l^2$ and  $\tilde \l=\l^2-1$.
The arrow point towards the IR and there is no flow in the white areas.
The physical region between the vertical lines at $\l=\pm 1 $ is mapped to the one outside by
\eqn{symmm}.}
\end{figure}

\no
We have checked that the more general three parameter case, does not
lead to new fixed points and that it also has regions where the $\b$-functions vanish identically.
We do not present any results since this case does not seem to give conceptually something new.

\no
Finally, note that for $\l=0$, the $\b$-functions vanish, implying that the theory stays scale invariant, even if we add the Lorentz-breaking  term. This is consistent with the fact that all current correlators do not develop logarithmic terms
at any order in perturbation theory. The latter is a statement can be readily checked.

\subsection{Vanishing of the $\b$-functions and the potentially  unstable region}

We will argue that the regions I and IV are  physically irrelevant when we consider fluctuations around the vacuum associated with the unit element $g=\mathbb{1}$, which obviously solves the equations of motion.
These regions are precisely those where the coupling $\l$ and $\tilde \l$ do not run at all.

\no
Expanding round the identity we have $g=1+i \,  X^a T^a$ and recalling that $\s^{\pm}=\tau \pm \s$, then
\eqn{defactigen} becomes
\be
\begin{split}
&\cL^{(0)}={k\ov 8 \pi}  \Big( (A+B) ( \del_\tau X)^2 -(A-B) ( \del_\s X)^2 -2 B \,  \del_\tau X \, \cdot\del_\s X \Big)\ ,
\\
&A={1+\l\ov1-\l}\ ,\qq B={ \tilde \l \ov (1-\l)^2}\ ,
\label{ggX}
\end{split}
\ee
where for simplicity we have suppressed the target space indices.
The Hamiltonian density reads
\be
\begin{split}
{\cal H}= {k\ov 8 \pi}  \Big( (A+B) ( \del_\tau X)^2 +(A-B) ( \del_\s X)^2  \Big)\ .
\label{Ham-den}
\end{split}
\ee
The equations of motion following from the Lagrangian \eqn{ggX} take the form
\be
\begin{split}\label{eom}
(A+B) \del^2_\tau X^a-(A-B) \del^2_\s X^a-2B \, \del_\tau \del_\s X^a=0\ .
\end{split}
\ee
This can be easily solved by making a plane wave ansatz of the form $X^a= C^a \sin{(\om \tau+n \s)}$, where $n\in Z$ is the winding number and the  frequency $\om$ is given by
\be
\om=n\quad {\rm or}\quad \om={B-A \ov A+B}n\ .
\ee
Substituting these values in \eqn{Ham-den} and integrating over $\s\!\in\!(0, 2 \pi)$, we get for the string
energy that
\be
\begin{split}\label{energy}
E_n={k\ov 4} A n^2\ , \qq E_n  ={k\ov 4} {A(A-B) \ov A+B} n^2 \ ,
\end{split}
\ee
for the two frequencies, respectively.
Hence, the string spectrum is unbounded from below whenever $A$ and/or the ratio ${A-B\ov A+B}$ becomes negative
(recall that $k$ is positive). By inspection it is easy to see that this happens when the couplings $\l$ and $\tilde \l$ lie in the regions I and IV or for the regions II and III the parameter $|\l|>1$ (see Figure 1).
Finally, It would be interesting to clarify if there exists some other vacuum around which the fluctuations have a positive definite spectrum.

\section{Discussion and future directions}

In this work, we have constructed certain two-dimensional $\s$-models in which Lorentz invariance is broken.
We have identified an integrable sector
in which we computed the RG $\b$-function equations for the couplings. The flow is non-trivial precisely
 in the region where the Hamiltonian is bounded. We found no new RG fixed points in the IR which is similar to the case of the prototype Lorentz preserving $\l$-deformations.
 For this to happen, experience shows that two WZW actions at different levels
should be involved similarly to the Lorentz invariant $\l$-deformed models of \cite{Georgiou:2017jfi}. It will be interesting to perform research  in this direction and identify the CFTs at the IR fixed points.

\no
It would be certainly interesting to calculate the anomalous dimensions of current  and primary operators, as well as  three-point correlators involving currents and/or primary operators as exact functions of the deformation parameters.
This should be done along the lines of  \cite{Georgiou:2015nka,Georgiou:2016iom,Georgiou:2016zyo} for Lorentz preserving $\l$-models which heavily used symmetry and analyticity arguments, combined with minimal perturbative information.

\no
Finally, it should be interesting to extend the present work by constructing Lorentz-breaking $\l$-deformed models based on coset spaces along the lines of \cite{Sfetsos:2013wia,Hollowood:2014rla,Sfetsos:2017sep}. In that case the Lorentz-breaking terms should be described by parafermions \cite{Fateev:1985mm,Bardacki:1990wj}.

\subsection*{Acknowledgments}

The work of G.G. on this project has received funding from the Hellenic Foundation for Research and Innovation
(HFRI) and the General Secretariat for Research and Technology (GSRT), under grant
agreement No 15425.




\begin{thebibliography}{1}


 \bibitem{Sfetsos:2013wia}
  K.~Sfetsos, {\it Integrable interpolations: From exact CFTs to non-Abelian T-duals},\hfill\break
  Nucl. Phys. {\bf B880} (2014) 225, \href{http://arxiv.org/abs/arXiv:1312.4560}{arXiv:1312.4560 [hep-th]}.

\bibitem{Georgiou:2016urf}
  G.~Georgiou and K.~Sfetsos,
  {\it A new class of integrable deformations of CFTs}, \hfill\break
  JHEP {\bf 1703} (2017) 083,
  \href{https://arxiv.org/abs/1612.05012}{arXiv:1612.05012 [hep-th]}.


\bibitem{Georgiou:2017oly}
  G.~Georgiou, K.~Sfetsos and K.~Siampos,
  {\it Double and cyclic $\lambda$-deformations and their canonical equivalents},
  Phys. Lett. {\bf B771}, 576 (2017),
   \href{https://arxiv.org/abs/1704.07834}{arXiv:1704.07834 [hep-th]}.


  \bibitem{Georgiou:2017jfi}
  G.~Georgiou and K.~Sfetsos,
  {\it Integrable flows between exact CFTs},\\
  JHEP {\bf 1711}, 078 (2017),
   \href{https://arxiv.org/abs/1707.05149}{arXiv:1707.05149 [hep-th]}.


\bibitem{Sfetsos:2017sep}
  K.~Sfetsos and K.~Siampos,
  {\it Integrable deformations of the $G_{k_1} \times G_{k_2}/G_{k_1+k_2}$ coset CFTs},
  Nucl. Phys. {\bf B927}, 124 (2018),
  \href{https://arxiv.org/abs/1710.02515}{arXiv:1710.02515  [hep-th]}.


\bibitem{Georgiou:2018hpd}
  G.~Georgiou and K.~Sfetsos,
  {\it Novel all loop actions of interacting CFTs: Construction, integrability and RG flows},
  Nucl. Phys. {\bf B937} (2018) 371,
  \href{https://arxiv.org/abs/1809.03522}{arXiv:1809.03522 [hep-th]}.


\bibitem{Georgiou:2018gpe}
  G.~Georgiou and K.~Sfetsos,
  {\it The most general $\lambda$-deformation of CFTs and integrability},
  \href{https://arxiv.org/abs/1812.04033} {arXiv:1812.04033 [hep-th]}.

    \bibitem{Appadu:2015nfa}
  C. Appadu and T.J. Hollowood,
  {\it Beta function of k deformed ${\text AdS}_{5} \times S^5$ string theory},
  JHEP {\bf 1511} (2015) 095,
  \href{http://arxiv.org/abs/arXiv:1507.05420}{arXiv:1507.05420 [hep-th].}


      \bibitem{Georgiou:2017aei}
  G.~Georgiou, E.~Sagkrioti, K.~Sfetsos and K.~Siampos,
  {\it Quantum aspects of doubly deformed CFTs},
Nucl. Phys. {\bf B919} (2017) 504,
 \href{https://arxiv.org/abs/1703.00462}
  {arXiv:1703.00462 [hep-th]}.



\bibitem{Sagkrioti:2018rwg}
  E.~Sagkrioti, K.~Sfetsos and K.~Siampos,
  {\it RG flows for $\lambda$-deformed CFTs},\\
  Nucl.\ Phys.\ {\bf B930} (2018) 499,
    \href{https://arxiv.org/abs/1801.10174}{arXiv:1801.10174 [hep-th].}

\bibitem{LeClair:2001yp}
  A.~LeClair,
  {\it Chiral stabilization of the renormalization group for flavor and color anisotropic current interactions},
  Phys.\ Lett.\ {\bf B519} (2001) 183,
  \href{https://arxiv.org/abs/hep-th/0105092v2}{hep-th/0105092}.

\bibitem{Georgiou:2016zyo}
  G.~Georgiou, K.~Sfetsos and K.~Siampos,
  {\it $\lambda$-deformations of left-right asymmetric CFTs}, Nucl. Phys. {\bf B914} (2017) 623,
\href{https://arxiv.org/abs/1610.05314}{arXiv:1610.05314 [hep-th]}.



\bibitem{Guica:2017lia}
  M.~Guica,
  {\it An integrable Lorentz-breaking deformation of two-dimensional CFTs},
  SciPost Phys.\  {\bf 5}, no. 5, 048 (2018)
  \href{https://arxiv.org/abs/1710.08415}{arXiv:1710.08415 [hep-th]}.



\bibitem{Smirnov:2016lqw}
  F.~A.~Smirnov and A.~B.~Zamolodchikov,
  {\it On space of integrable quantum field theories},
  Nucl.\ Phys. {\bf B915}, 363 (2017)
 \href{https://arxiv.org/abs/1608.05499}{arXiv:1608.05499 [hep-th]}.

\bibitem{Cavaglia:2016oda}
   R.~Conti, S.~Negro and R.~Tateo,
  {\it  $T \bar{T}$-deformed 2D Quantum Field Theories},
  JHEP {\bf 1610}, 112 (2016)
  \href{https://arxiv.org/abs/1608.05534}{arXiv:1608.05534 [hep-th]}.

\bibitem{Bzowski:2018pcy}
  A.~Bzowski and M.~Guica,
  {\it The holographic interpretation of $J \bar T$-deformed CFTs},
  JHEP {\bf 1901}, 198 (2019)
  \href{https://arxiv.org/abs/1803.09753}{arXiv:1803.09753 [hep-th]}.

\bibitem{Chakraborty:2018vja}
  S.~Chakraborty, A.~Giveon and D.~Kutasov,
  {\it $ J\overline{T} $ deformed CFT$_{2}$ and string theory},
  JHEP {\bf 1810}, 057 (2018)
   \href{https://arxiv.org/abs/1806.09667}
  {arXiv:1806.09667 [hep-th]}.

\bibitem{Apolo:2018qpq} 
  L.~Apolo and W.~Song,
  {\it Strings on warped AdS$_{3}$ via $ \mathrm{T}\bar{\mathrm{J}} $ deformations},
  JHEP {\bf 1810}, 165 (2018)
   \href{https://arxiv.org/abs/1806.10127}
  {arXiv:1806.10127 [hep-th]}.

  
\bibitem{Aharony:2018ics}
  O.~Aharony, S.~Datta, A.~Giveon, Y.~Jiang and D.~Kutasov,
  {\it Modular covariance and uniqueness of $J\bar{T}$ deformed CFTs},
  JHEP {\bf 1901}, 085 (2019)doi:10.1007/JHEP01(2019)085
  \href{https://arxiv.org/abs/1808.08978}
  {arXiv:1808.08978 [hep-th]}.

   \bibitem{Georgiou:2015nka}
  G.~Georgiou, K.~Sfetsos and K.~Siampos,
  {\it All-loop anomalous dimensions in integrable $\lambda$-deformed $\sigma$-models},
  Nucl.\ Phys.\  {\bf B901} (2015) 40,
  \href{http://arxiv.org/abs/1509.02946}{arXiv:1509.02946 [hep-th].}


\bibitem{Georgiou:2016iom}
  G.~Georgiou, K.~Sfetsos and K.~Siampos,
  {\it All-loop correlators of integrable $\l$-deformed $\s$-models},
  Nucl.  Phys. {\bf B909} (2016) 360,
  \href{http://arxiv.org/abs/arXiv:1604.08212}{1604.08212 [hep-th].}


    \bibitem{Itsios:2014lca}
  G.~Itsios, K.~Sfetsos and K.~Siampos,
  {\it The all-loop non-Abelian Thirring model and its RG flow},
  Phys.\ Lett.\  {\bf B733} (2014) 265,
  \href{http://arxiv.org/abs/1404.3748}{arXiv:1404.3748 [hep-th].}

 \bibitem{Sfetsos:2014jfa}
  K.~Sfetsos and K.~Siampos,
  {\it Gauged WZW-type theories and the all-loop anisotropic non-Abelian Thirring model},
  Nucl. Phys.  {\bf B885} (2014) 583,
  \href{http://arxiv.org/abs/arXiv:1405.7803}{arXiv:1405.7803 [hep-th].}


  \bibitem{Kutasov:1989aw}
  D.~Kutasov, {\it Duality Off the Critical Point in Two-dimensional Systems With Nonabelian Symmetries},
\href{http://www.sciencedirect.com/science/article/pii/0370269389913257}{Phys. Lett. {\bf B233} (1989) 369}.

\bibitem{Guica:2019vnb}
  M.~Guica,
  {\it On correlation functions in $J\bar T$-deformed CFTs},
   \href{https://arxiv.org/abs/1902.01434}{arXiv:1902.01434 [hep-th]}.

     \bibitem{Nappi:1979ig}
  C.R. Nappi,
  {\it Some Properties of an Analog of the Nonlinear $\sigma$-Model},\\
 \href{ http://journals.aps.org/prd/abstract/10.1103/PhysRevD.21.418}
  {Phys. Rev. {\bf D21} (1980) 418}.


  \bibitem{honer}
 G.~Ecker and J.~Honerkamp,
 {\it Application of invariant renormalization to the nonlinear chiral invariant
 pion Lagrangian in the one-loop approximation},\hfill\break
 \href{http://www.sciencedirect.com/science/article/pii/0550321371904688}{Nucl. Phys. {\bf B35} (1971) 481}.\hfill\break
J.~Honerkamp,
 {\it Chiral multiloops},
\href{http://www.sciencedirect.com/science/article/pii/0550321372902994}{Nucl. Phys. {\bf B36} (1972) 130}. \hfill\break
  D.~Friedan,
  {\it Nonlinear Models in Two Epsilon Dimensions},\hfill\break
  \href{http://journals.aps.org/prl/abstract/10.1103/PhysRevLett.45.1057}{Phys. Rev. Lett. {\bf 45} (1980) 1057}
 and {\it Nonlinear Models in Two + Epsilon Dimensions},
  \href{http://www.sciencedirect.com/science/article/pii/0003491685903847}{Annals Phys. {\bf 163} (1985) 318}.


    \bibitem{Hollowood:2014rla}
  T.J.~Hollowood, J.L.~Miramontes and D.M.~Schmidtt,
 {\it Integrable Deformations of Strings on Symmetric Spaces},
  JHEP {\bf 1411} (2014) 009,
  \href{http://arxiv.org/abs/1407.2840}{arXiv:1407.2840 [hep-th]}.


\bibitem{Fateev:1985mm}
  V.A.~Fateev and A.B.~Zamolodchikov,
  {\it Parafermionic Currents in the Two-Dimensional Conformal Quantum Field Theory
and Selfdual Critical Points in Z(n) Invariant Statistical Systems},\hfill\break
  Sov. Phys. JETP {\bf 62} (1985) 215  [Zh. Eksp. Teor. Fiz.  {\bf 89} (1985) 380].

\bibitem{Bardacki:1990wj}
  K.~Bardacki, M.J.~Crescimanno and E.~Rabinovici,
  {\it Parafermions from Coset Models},
 \href{http://www.sciencedirect.com/science/article/pii/055032139190332R}{Nucl. Phys. {\bf B344} (1990) 344}.
 \hfill\break
  K.~Bardakci, M.J.~Crescimanno and S.~Hotes,
  {\it Parafermions from nonabelian coset models},
 \href{http://www.sciencedirect.com/science/article/pii/055032139190332R}{Nucl. Phys. {\bf B349} (1991) 439}.


\end{thebibliography}
\end{document}